\documentclass[fleqn,10pt]{wlscirep}
\usepackage[utf8]{inputenc}
\usepackage[T1]{fontenc}
\usepackage{hyperref}
\usepackage[final]{pdfpages}
\makeatletter
\AtBeginDocument{\let\LS@rot\@undefined}
\makeatother

\title{Universality of political corruption networks}
\author[1]{Alvaro F. Martins}
\author[2,3]{Bruno R. da Cunha}
\author[4]{Quentin S. Hanley}
\author[5]{Sebasti\'an Gon\c{c}alves}
\author[6,7,8,9,*]{Matja{\v z} Perc}
\author[1,$\dagger$]{Haroldo V. Ribeiro}

\affil[1]{Departamento de F\'isica, Universidade Estadual de Maring\'a -- Maring\'a, PR 87020-900, Brazil}

\affil[2]{Rio Grande do Sul Superintendency, Brazilian Federal Police -- Porto Alegre, RS 90160-093, Brazil}
\affil[3]{National Police Academy, Brazilian Federal Police -- Brasília, DF 71559-900, Brazil}

\affil[4]{School of Science and Technology, Nottingham Trent University, Clifton Lane, Nottingham NG11 8NS, United Kingdom}

\affil[5]{Instituto de F\'isica, Universidade Federal do Rio Grande do Sul -- Porto Alegre, RS 91501-970, Brazil}

\affil[6]{Faculty of Natural Sciences and Mathematics, University of Maribor, Koro{\v s}ka cesta 160, 2000 Maribor, Slovenia}
\affil[7]{Department of Medical Research, China Medical University Hospital, China Medical University, Taichung, Taiwan}
\affil[8]{Alma Mater Europaea, Slovenska ulica 17, 2000 Maribor, Slovenia}
\affil[9]{Complexity Science Hub Vienna, Josefst{\"a}dterstra{\ss}e 39, 1080 Vienna, Austria}

\affil[*]{email: matjaz.perc@gmail.com}
\affil[$\dagger$]{email: hvr@dfi.uem.br}

\begin{abstract}
Corruption crimes demand highly coordinated actions among criminal agents to succeed. But research dedicated to corruption networks is still in its infancy and indeed little is known about the properties of these networks. Here we present a comprehensive investigation of corruption networks related to political scandals in Spain and Brazil over nearly three decades. We show that corruption networks of both countries share universal structural and dynamical properties, including similar degree distributions, clustering and assortativity coefficients, modular structure, and a growth process that is marked by the coalescence of network components due to a few recidivist criminals. We propose a simple model that not only reproduces these empirical properties but reveals also that corruption networks operate near a critical recidivism rate below which the network is entirely fragmented and above which it is overly connected. Our research thus indicates that actions focused on decreasing corruption recidivism may substantially mitigate this type of organized crime.
\end{abstract}

\begin{document}

\flushbottom
\maketitle

\thispagestyle{empty}

\section*{Introduction}

Understanding the collective and intricate nature of political corruption and other organized crime demands more than simple statistics. In an analogy with complex systems~\cite{jensen1998self, mitchell2009complexity, castellano2009statistical, jusup2021social}, where the whole is often more than just the sum of its parts, one may say that the success of criminal organizations depends not only on the individual skills of the criminals, but in fact much more so on their ability to cooperate and create robust organizational structures that are capable of protecting and hiding their illegal activities. The usage of complexity science has been advocated by different authors as an ideal framework to investigate economic crime, organized crime, and corruption~\cite{d2015statistical, luna2020corruption, kertesz2021complexity, granados2021corruption, da2021criminofisica}. Network science~\cite{newman2010networks, barabasi2015netork} stands out in this context as it can most suitably describe the different interactions among criminals by means of a wide array of tools and methods that have been developed over the past two decades~\cite{luna2020corruption}. 

Empirical investigations of criminal networks are, however, often made difficult by the unavailability of reliable data about these systems, especially time-resolved data. This is in part because criminals do their best to remain undetected, but also, and in fact primarily so, because this information is often classified and restricted to law enforcement agencies. Despite these difficulties, many recent works have demonstrated the usefulness of network science to investigate criminal networks, with examples including cartel detection~\cite{wachs2019network}, corruption risk in contracting markets~\cite{wachs2021corruption}, money laundering~\cite{garcia2020ai}, identification of corrupt politicians via voting networks~\cite{colliri2019analyzing}, dark web pedophile rings~\cite{da2020assessing}, criminal conspiracy networks of companies~\cite{nicolas2021conspiracy}, modular structure of crime organizations~\cite{calderoni2017communities}, political corruption networks~\cite{ribeiro2018dynamical}, organized crime networks~\cite{joseph2021ties}, controllability of criminal networks~\cite{solimine2020political}, resilience of drug trafficking~\cite{duijn2014relative}, as well as police criminal intelligence networks~\cite{da2018topology}. Nevertheless, and despite the fascinating research that has already been made, we still need to identify overarching common properties and dynamical aspects of criminal networks, which might allow us to develop simple models that describe fundamental features and provide valuable insights into organized crime. 

Here we aim to address this challenge by presenting a comprehensive investigation of static and dynamical aspects of political corruption networks that are associated with corruption scandals in Brazil and Spain. The two datasets that we use provide unprecedented information about corruption activities spanning almost three decades of history in each country and involve over 400 people in Brazil and more than 2,700 people in Spain. Our research shows that corruption scandals in both countries rarely involve more than ten people, and that the corresponding corruption networks have universal properties such as an exponential degree distribution, a high clustering coefficient, the small-world property, homophily, modular structure, and a similar relation between the number of network modules and the total number of political scandals. We also study the temporal aspects of these scandals to create an evolving network representation, which allows us to identify striking similarities in the dynamics of political corruption networks of both countries. Specifically, we find that the giant component of these networks exhibits abrupt changes caused by the coalescence of network components, that the number of network modules increases linearly with the number of scandals, and that the number of recidivist agents (people involved in more than one corruption scandal) is linearly associated with the number of people in the network over the entire network growth process. These empirical universalities finally allow us to propose a simple model where the main parameter is the recidivism rate, that is, the fraction of corrupt agents recurring in the criminal activity. Beyond simulating corruption networks with features very similar to those observed in the empirical data, our model indicates that corruption networks operate around a critical recidivism rate below which the network becomes completely fragmented and above which it is overly connected.

\section*{Results}

We start by presenting the two datasets of corruption scandals used in our study. The Brazilian data is the same reported in Ref.~\cite{ribeiro2018dynamical} and comprises 65 well-documented political corruption scandals that occurred in Brazil between 1987 and 2014. This information was manually compiled from web pages of magazines and newspapers with wide circulation and includes the names of the 404 people involved in each of the 65 scandals. The Spanish data are original to our work and have been extracted in May 2020 from a non-profit website~\cite{casosaislados} that aims to list all known corruption scandals in Spain. The information on this website is also compiled from publicly accessible web pages of popular Spanish news magazines and daily newspapers. The Spanish data comprises 437 corruption scandals that occurred between 1989 and 2018 and involved 2,753 people. 

\begin{figure*}[!ht]
  \centering
  \includegraphics[width=0.95\textwidth, keepaspectratio]{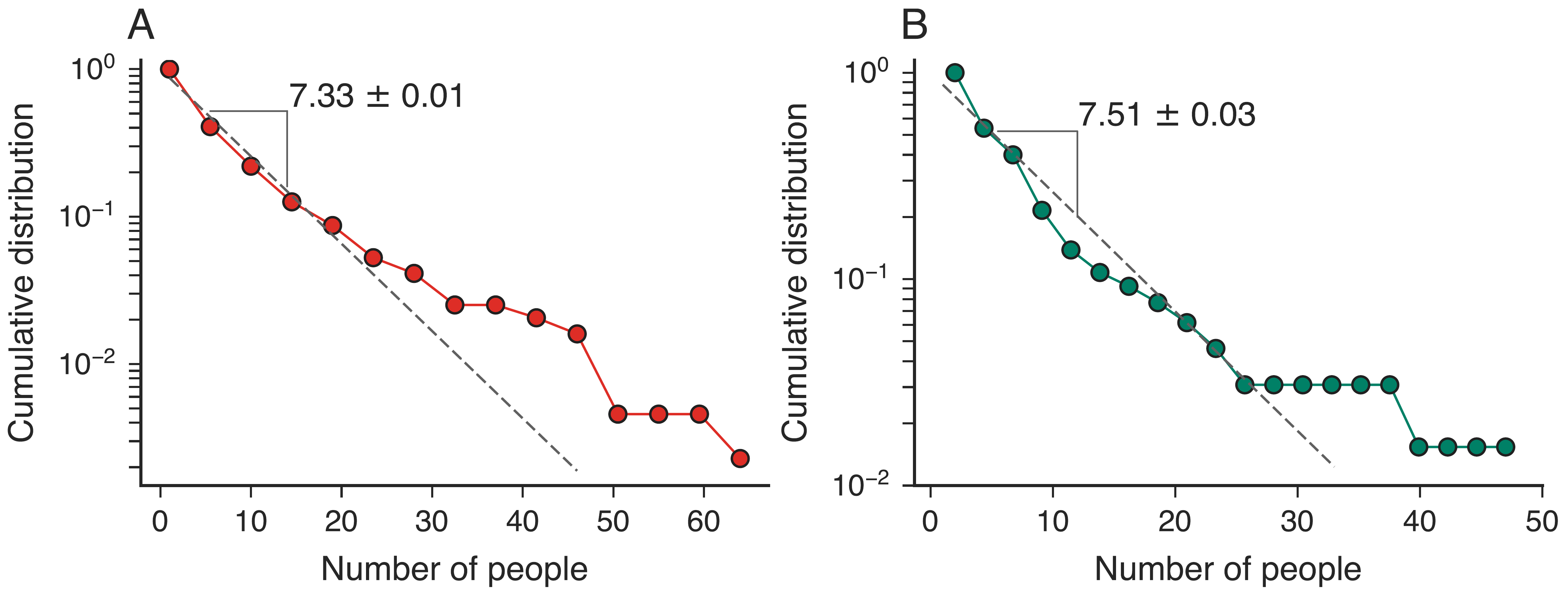}
  \caption{The size of corruption scandals is approximately exponentially distributed. Complementary cumulative distribution function of the number of people involved in political corruption scandals in (A) Spain and (B) Brazil. The dashed lines indicate an exponential distribution adjusted to data via the maximum likelihood method. The characteristic number of people involved in these scandals (indicated by the numbers within each panel) is around seven people for both countries.}
  \label{fig:1}
\end{figure*}

Having described our datasets, we first examine the distribution associated with the size (number of people involved) of corruption scandals. As reported in Ref.~\cite{ribeiro2018dynamical} for Brazil, we find that the size distribution of scandals is roughly approximated by an exponential distribution with a characteristic number of people around seven people for both countries (Figs.~\ref{fig:1}A and~\ref{fig:1}B). Despite the deviations between the exponential model and the empirical distributions observed for large scandals, this result shows that political corruption runs in small groups that rarely exceed more than ten people (only 20\% and 17\% of corruption cases in Spain and Brazil, respectively). Thus, it seems that corrupt agents usually rely on a small number of cronies for running their criminal activities, probably because large-scale processes are hard to manage and remain undetected for longer periods~\cite{baker1993social}. Moreover, the surprising similarities in size distributions of scandals in both countries already indicate a possible universal pattern related to political corruption processes.

\begin{figure*}[!ht]
  \centering
  \includegraphics[width=1\textwidth, keepaspectratio]{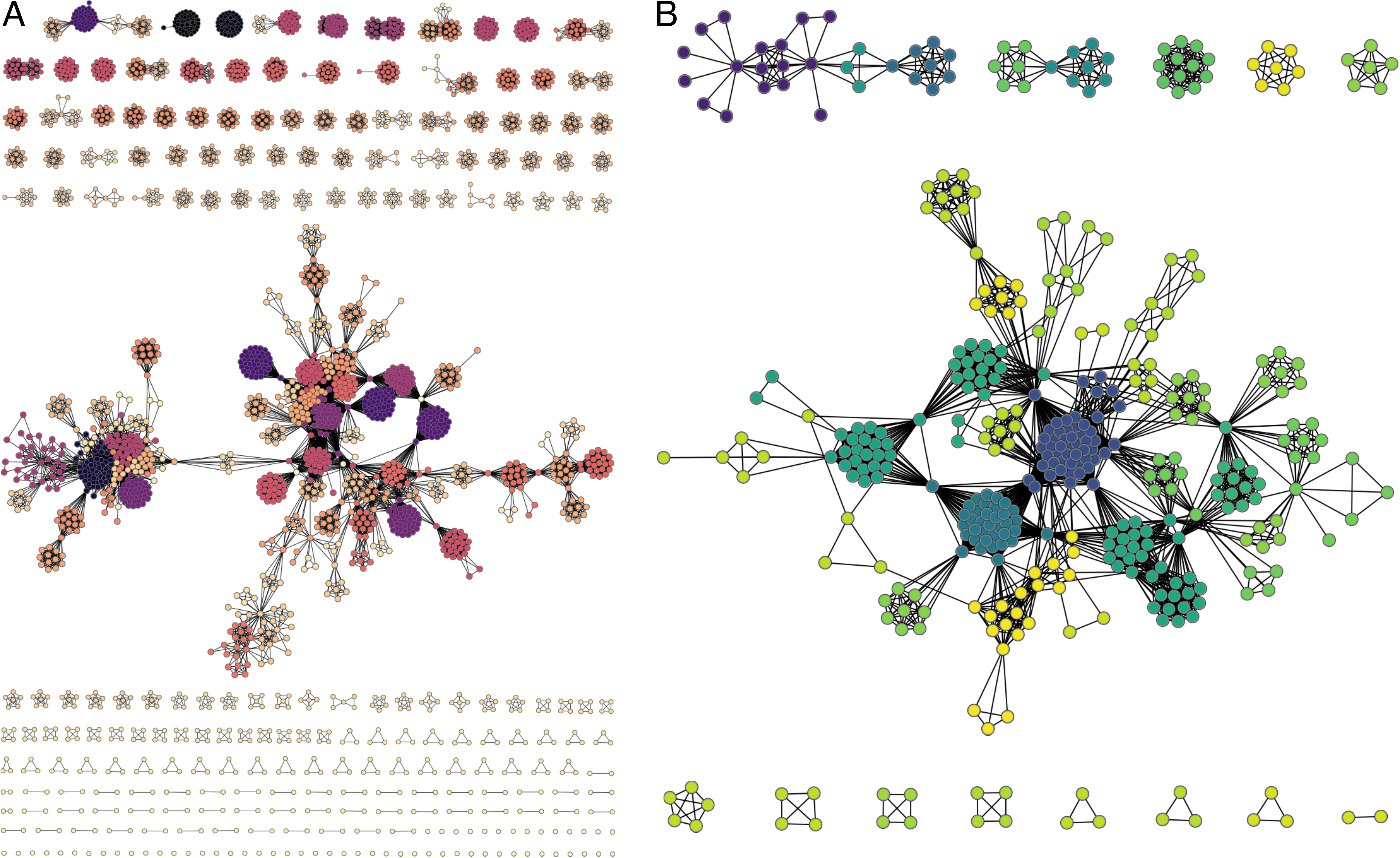}
  \caption{Visualization of the corruption networks formed by people involved in political scandals in (A) Spain and (B) Brazil. In both networks, nodes represent people and the edges among them indicate individuals engaged in the same corruption case. The colors refer to the modular structures of these networks estimated by the infomap algorithm~\cite{rosvall2008maps, rosvall2009map}.}
  \label{fig:2}
\end{figure*}

To investigate the emerging patterns of people involved in corruption cases, we have created a static network representation of these scandals where people are nodes and connections among them indicate individuals engaged in the same corruption case. Figures~\ref{fig:2}A and \ref{fig:2}B depict the Spanish and Brazilian corruption networks, respectively. The Spanish network has 2,753 nodes, 27,545 edges, 197 connected components, 58 isolated nodes, and a giant component accounting for 40\% of nodes and 53\% of edges. In turn, the Brazilian network comprises 404 nodes, 3,549 edges, 14 connected components, and a giant component accounting for 77\% of nodes and 93\% of edges. 

\begin{figure*}[!t]
  \centering
  \includegraphics[width=0.95\textwidth, keepaspectratio]{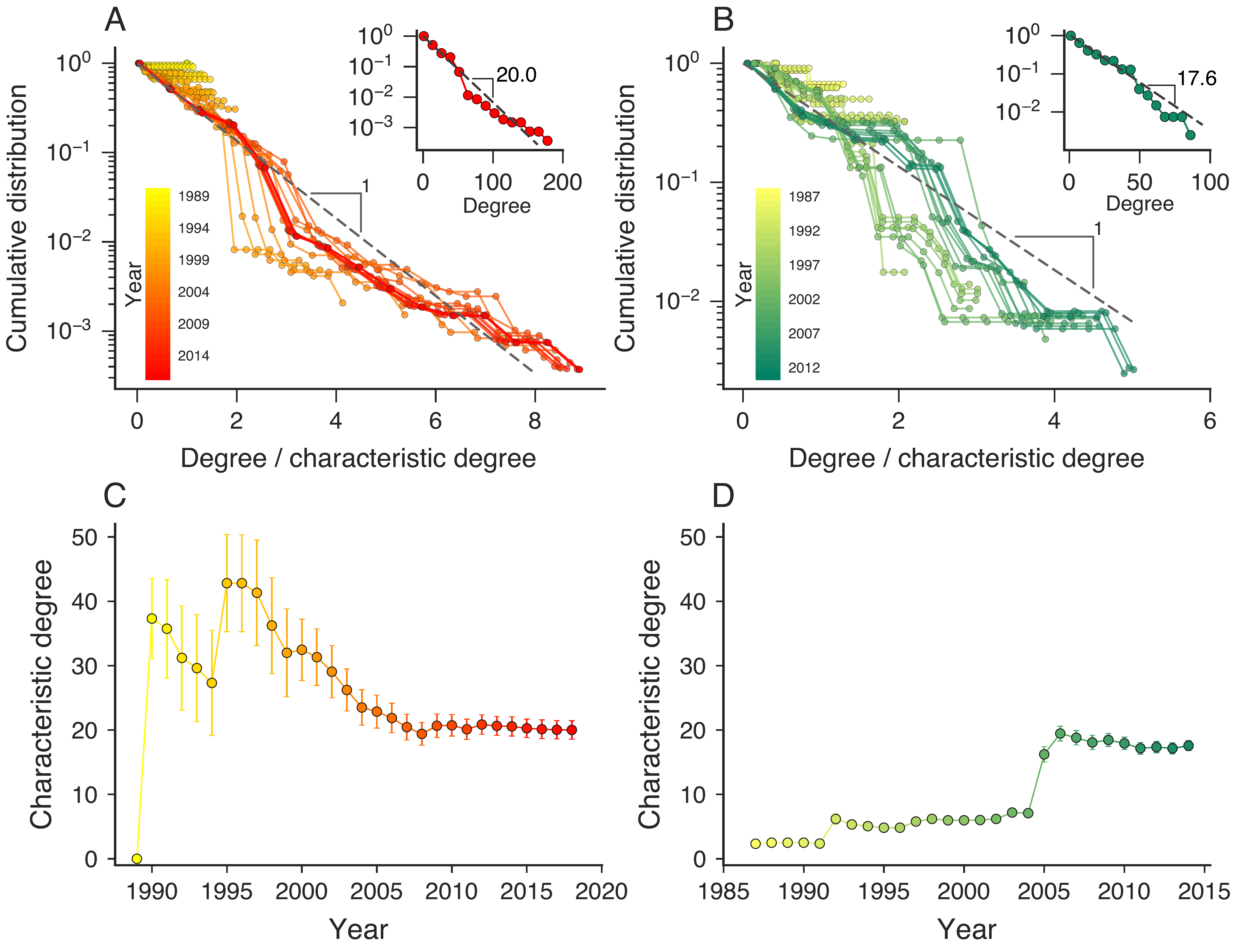}
  \caption{Degree distributions of corruption networks are approximated by exponential models with characteristic degrees that seem to approach a constant value with the network growth. Complementary cumulative distributions of the node degree divided by the characteristic degree for each year (indicated by the color code) of the (A) Spanish and (B) Brazilian networks. The insets show the degree distributions for the latest stage of the network of each country. The approximately linear behavior of these curves on a log-linear scale and the good quality collapse of the distributions indicate that the exponential model approximates well the degree distributions. Evolution of the characteristic degree for the (C) Spanish and (D) Brazilian corruption networks. The markers indicate the maximum-likelihood estimate of the characteristic degree in each year and the error bars stand for 95\% bootstrap confidence intervals.}
  \label{fig:3}
\end{figure*}

Despite the different sizes, these networks share striking similarities. For instance, both networks (Spanish vs. Brazilian networks, respectively) have high clustering coefficients (0.91 vs. 0.93 for the entire network and 0.94 vs. 0.93 for the giant component), moderately high assortativity coefficients (0.74 vs. 0.53 for the entire network and 0.59 vs. 0.50 for the giant component), low densities (0.007 vs. 0.044 for the entire network and 0.025 vs. 0.069 for the giant component), and small average shortest path length (5.11 vs. 2.99 for the giant component). Modular structures also characterize these networks (see colors in Figs.~\ref{fig:2}A and \ref{fig:2}B) and usually merge more than one scandal into a single network module, as estimated by the infomap clustering algorithm~\cite{rosvall2008maps, rosvall2009map}. Indeed, the ratio between modules and scandals is 0.76 for the Spanish network and 0.62 for the Brazilian one. Moreover, we find that the degree distributions of both networks are well-approximated by exponential distributions with characteristic degrees equal to 20.0 people for Spain and 17.6 people for Brazil (see insets of Figs.~\ref{fig:3}A and \ref{fig:3}B).

Beyond the previous static representation, our data allow us to investigate dynamical patterns associated with the growth of these corruption networks over time. To do so, we create time-dependent networks of people involved in corruption scandals up to a given year. Then, by increasing this threshold year, we observe a process of network growth in which new nodes and edges among new and old nodes emerge year after year due to the discovery of new scandals. Using this time-varying representation, we first ask whether the approximated exponential degree distribution holds for all years. We have fitted the exponential model to each stage of our networks via the maximum-likelihood method, and the results indicate that the degree distributions are in good agreement with the exponential distribution for all years of both Spanish and Brazilian networks. Figures~\ref{fig:3}A and \ref{fig:3}B show the complementary cumulative degree distributions divided by the characteristic degree (insets depict the degree distributions for the latest network stage), where the linear behavior on the log-linear scale and the good quality collapse of the distributions support the exponential hypothesis. Moreover, Figs.~\ref{fig:3}C and \ref{fig:3}D depict the evolution of the characteristic degree of our corruption networks. We observe significant variations in earlier network stages followed by an approximately steady characteristic degree in later stages that is surprisingly similar for the two countries.

\begin{figure*}[!ht]
  \centering
  \includegraphics[width=0.95\textwidth, keepaspectratio]{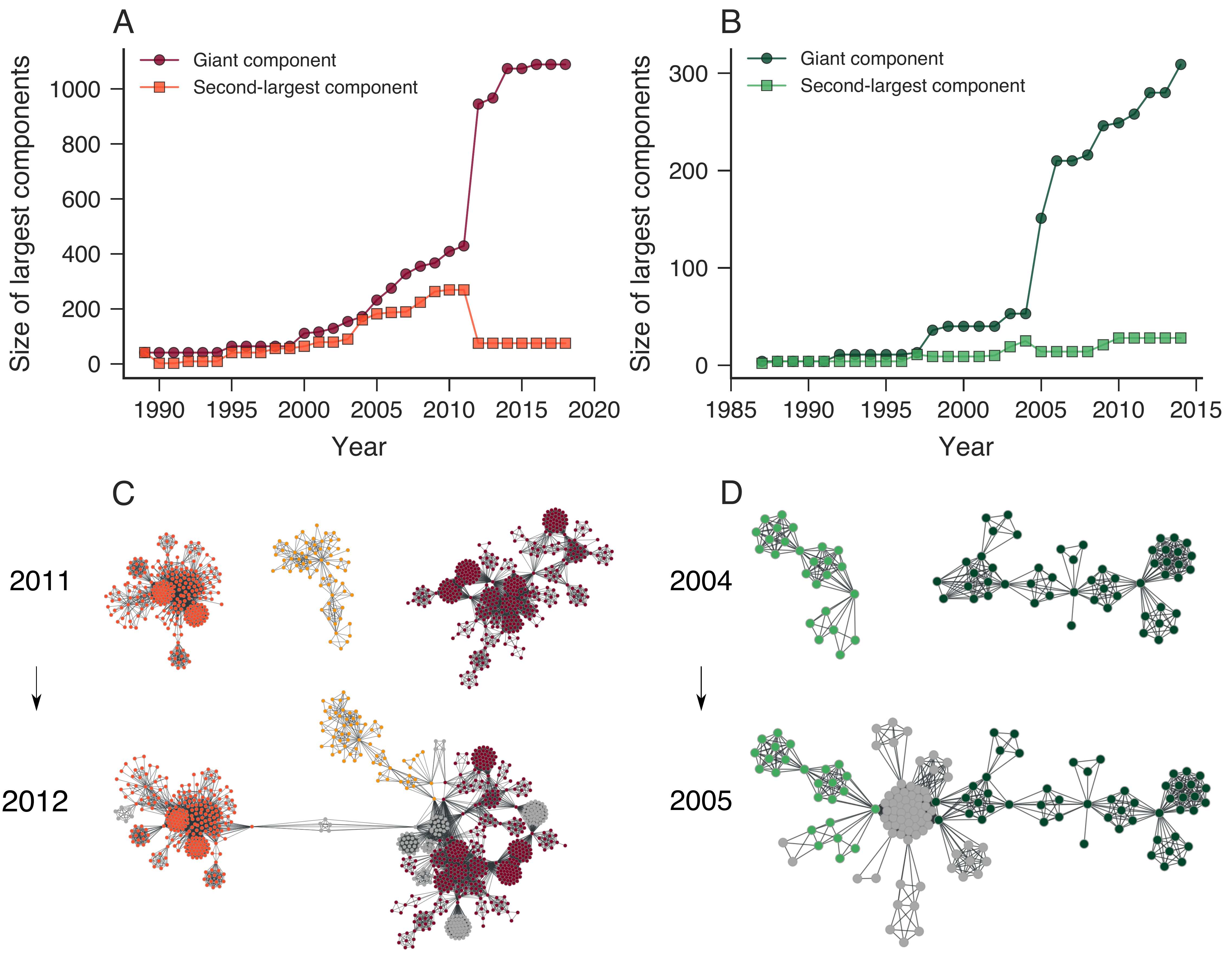}
  \caption{Corruption networks grow by a coalescence-like process of network components. Evolution of the size of the giant component (circle) and second-largest component (square) of the (A) Spanish and (B) Brazilian corruption networks. We observe that these quantities undergo sudden changes between specific years (2011–2012 for Spain and 2004–2005 for Brazil) characterized by an abrupt increase in the giant component and an abrupt decrease in the second-largest component. Snapshot visualizations of the network before and after the abrupt changes in the largest components of the (C) Spanish and (D) Brazilian networks. We observe that these changes are associated with a coalescence of network components caused by the emergence of new scandals (new nodes are colored in gray) involving a few recidivist agents.}
  \label{fig:4}
\end{figure*}

We have also investigated how the size of the main components of our corruption networks changes over time. Figures~\ref{fig:4}A and Figure~\ref{fig:4}B show the evolution of these quantities for the giant and second-largest components, where we find abrupt changes between particular years. For the Spanish network, the giant component steeply increases between 2011 and 2012, while the second-largest component abruptly shrinks during the same time interval. The Brazilian network exhibits similar patterns between the years 2004 and 2005, as also reported in Ref.~\cite{ribeiro2018dynamical}. This behavior is qualitatively similar to what happens in percolation transitions~\cite{bunde2012fractals} and indicates the existence of a coalescence-like process of network components. Indeed, by visualizing snapshots of our corruption networks (Figs.~\ref{fig:4}C and \ref{fig:4}D), we discover that the emergence of new political scandals involving a few recidivist agents causes the abrupt changes observed in the largest components.

\begin{figure*}[!ht]
  \centering
  \includegraphics[width=0.95\textwidth, keepaspectratio]{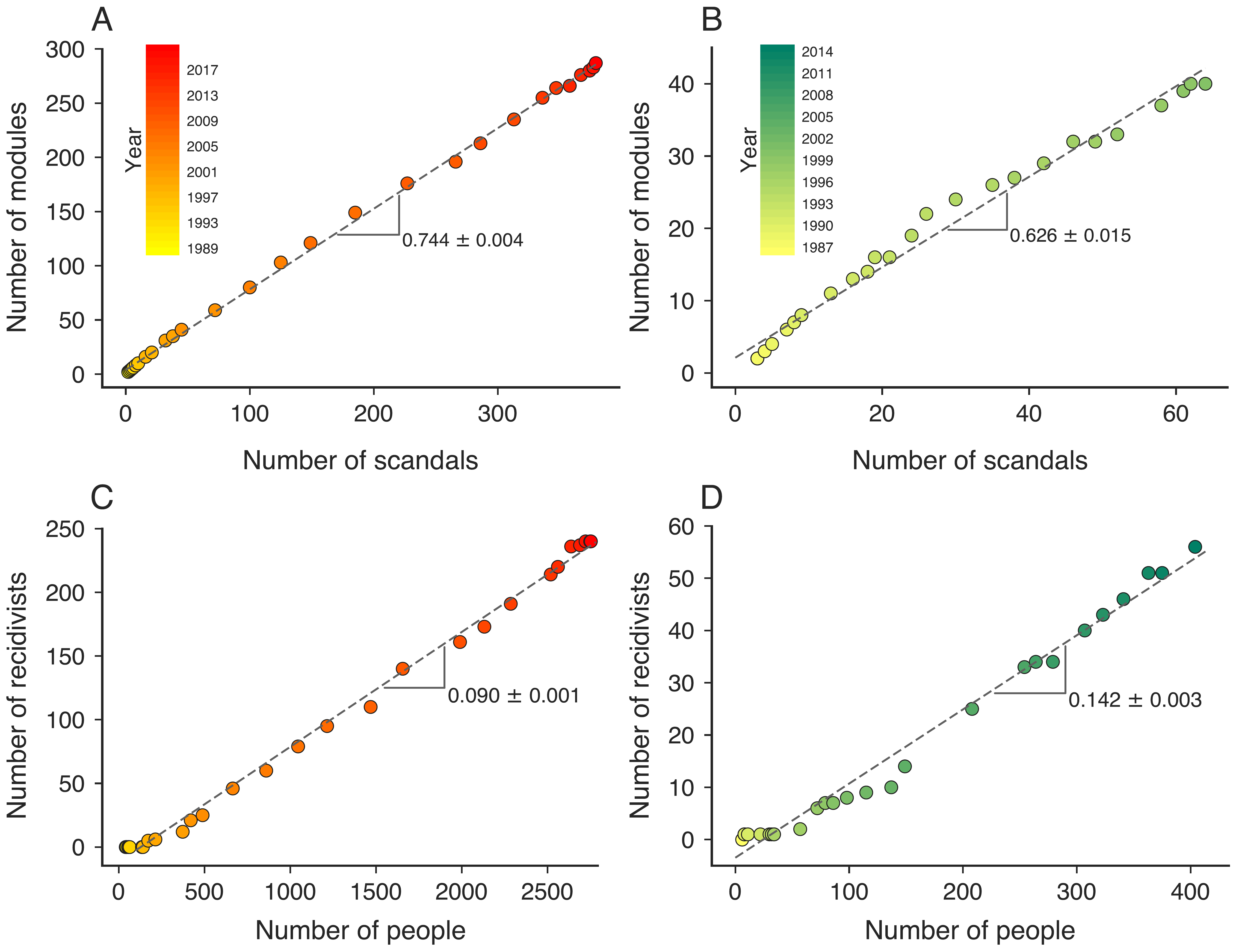}
  \caption{Evolution of the modular structure and the emergence of recidivist agents in corruption networks. Relationship between the number of network modules and the total of political scandals for each year of the (A) Spanish and (B) Brazilian corruption networks. The dashed lines are linear models adjusted to data, indicating that $0.744 \pm 0.004$ network modules are created per scandal in the Spanish network, while $0.626 \pm 0.015$ network modules per scandal emerge in the Brazilian network. Association between the number of recidivist agents and the total number of people for each year of the (C) Spanish and (D) Brazilian corruption networks. The dashed lines represent a linear model adjusted to data, where we find $0.090 \pm 0.001$ and $0.142 \pm 0.003$ recidivists per agent in the Spanish and Brazilian networks, respectively.}
  \label{fig:5}
\end{figure*}

As we have already shown, the latest stage of these networks displays a modular structure in which two or more scandals are usually merged into a single network module (Fig.~\ref{fig:2}). We now ask whether this behavior is particular to later network stages or a more general property of different stages of corruption networks. To answer this question, we examine the modular structure of these networks year after year using the infomap algorithm~\cite{rosvall2008maps, rosvall2009map} and determine the association between the number of network modules and the total of corruption scandals. While there is no fail-safe method for community or modular structure detection in networks~\cite{fortunato2010community, fortunato2016community}, we use the infomap due to its computational efficiency and good performance in benchmark tests with planted partition models~\cite{fortunato2010community, fortunato2016community}; however, we find similar results with modularity maximization or stochastic block models. Figures~\ref{fig:5}A and \ref{fig:5}B show that the number of network modules grows linearly with the total of political scandals with similar rates for both countries (0.744 modules per scandal for Spain 0.626 modules per scandal for Brazil). Thus, despite the underlying complexity of corruption processes, the structure of corruption networks approximately preserves the ratio between number of modules and scandals over their entire growth process. It is also worth remarking that this precise balance between modules and scandals is driven by the emergence of recidivist agents responsible for connecting different political scandals.

The dynamics of the largest network components and the linear association between modules and scandals exposed the critical role recidivist agents have on the structure of these corruption networks. To further understand the emergence of these special agents, we have investigated how the number of recidivist agents increases as new scandals are discovered and added to our corruption networks. Figures~\ref{fig:5}C and \ref{fig:5}D show the relation between the number of recidivist agents and the total of people for each year of the corruption networks of both countries. We observe that these two quantities are linearly associated, which implies that agents become recidivists at an approximately constant rate over the years. By fitting a linear model to the association between the number of recidivist agents and the total of people, we find the recidivism rate to be $0.090 \pm 0.001$ recidivists per agent for Spain and $0.142 \pm 0.003$ recidivists per agent for Brazil. These rates indicate that we expect to find about nine recidivists every hundred corrupt agents in the Spanish network. In comparison, the Brazilian network has about fourteen recidivists per hundred corrupt agents. Moreover, the higher recidivism rate observed for Brazil partially explains why the Brazilian network is denser and characterized by a lower average shortest path length than the Spanish counterpart.

Motivated by our empirical findings and the commonalities between the Spanish and Brazilian networks, we propose a simple model describing these corruption networks. This model starts with an empty network that grows by including complete graphs representing political scandals at each iteration. The number of people or the size of these complete graphs ($s$) is randomly drawn from an exponential distribution ($P$) to mimic the empirical behavior (Fig.~\ref{fig:1}), that is, $P(s)\sim e^{-s/s_c}$, where $s_c$ represents the characteristic size of corruption scandals (empirically, $s_c\approx 7$ people). We consider that part of the agents added to the network at each iteration are recidivists. By following the empirical behavior (Figs.~\ref{fig:5}C and \ref{fig:5}D), we assume the number of recidivists ($r$) to increase linearly with the total number of agents ($n$) via $r = \alpha n - \beta$, where $\alpha$ is the recidivism rate and $\beta>0$ controls the minimal number of people necessary for the emergence of the first recidivist agents. We keep track of the number of recidivists during the network growth process, and when new recidivists emerge, we randomly select nodes already present in the network to become recidivists and make them belong to the next scandal (complete graph) added to the network. Moreover, when selecting nodes for representing recidivist agents, we can select nodes that were already recidivists with a small probability $p$ or nodes that will become recidivists for the first time with probability $1-p$. This last procedure allows us to control the number of agents involved in more than two corruption scandals and reproduce the empirical behavior as about 2.5\% of all agents of both Spanish and Brazilian networks fit this condition.

\begin{figure*}[!ht]
  \centering
  \includegraphics[width=0.95\textwidth, keepaspectratio]{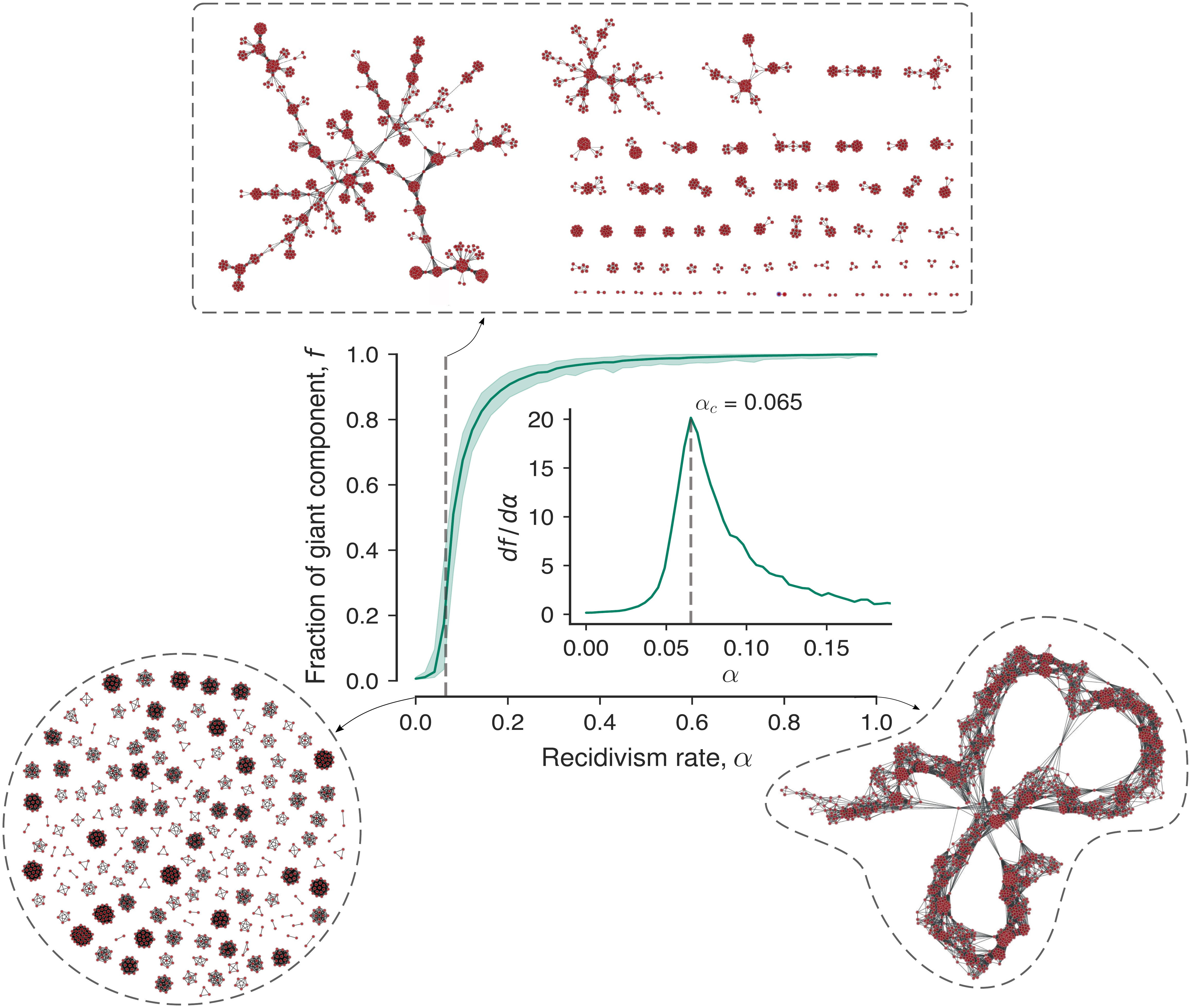}
  \caption{Corruption networks seem to operate close to the critical recidivism rate of our model. The continuous curve in the main panel shows the average fraction of the giant component of simulated networks ($f$) as a function of the recidivism rate ($\alpha$), and the shaded region represents the minimum and maximum values of $f$ estimated from one thousand model realizations for each $\alpha$. The inset in the main panel depicts the derivative of $f$ with respect to $\alpha$, and the dashed vertical line (also shown in the main panel) indicates the critical recidivism rate $\alpha_c = 0.065$ of our model, a value that is not too far from the recidivism rates of the Spanish ($\alpha=0.09$) and Brazilian  ($\alpha=0.14$) networks. The three visualizations surrounded by dashed paths represent typical simulated networks for $\alpha = 0$, $\alpha = \alpha_c$, and $\alpha = 1$. We observe that $\alpha = \alpha_c$ (upper network visualization) generates networks visually similar to the empirical corruption networks.}
  \label{fig:6}
\end{figure*}

We have generated networks using this model for different parameters and observed that the recidivism rate $\alpha$ is the most relevant parameter for the network structure. Because of this, we have fixed $s_c=7$, $\beta = 12$, and $p = 0.025$ (values close to their empirical counterparts) and explored the network behavior for different values of $\alpha$. To do so, we grow one thousand networks with one thousand iterations each (that is, by adding 1,000 complete graphs) for 50 values of $\alpha$ uniformly distributed between 0 and 1. Using this ensemble of networks, we evaluate the average fraction of the giant component $f$ (the number of nodes belonging to the largest component divided by the total of nodes) at the latest network stage as a function of the recidivism rate $\alpha$. Figure~\ref{fig:6} shows this analysis and three examples of simulated networks for $\alpha = 0$, $\alpha = 0.065$ and $\alpha = 1$. We note that small values of $\alpha$ produce very fragmented networks with giant components comprising only a small fraction of the network nodes. Conversely, values of $\alpha$ close to $1$ generate very connected networks with a chain-like structure and a giant component incorporating almost all network nodes. Similar to what happens in percolation transitions, we observe that the fraction of the giant component ($f$) steeply increases between $\alpha=0$ and $\alpha=0.1$. By calculating the derivative of the association between $f$ and $\alpha$ (inset of Fig.~\ref{fig:6}), we find a distinct peak that defines a critical recidivism rate $\alpha_c=0.065$ capable of generating networks visually similar to the empirical corruption networks (see the network example in Fig.~\ref{fig:6}). Interestingly, this critical recidivism rate is close to the empirical rates estimated for the Spanish ($\alpha=0.09$) and Brazilian ($\alpha=0.14$) networks. We have also investigated the model behavior near this critical point and the results suggest that this transition from very fragmented to chain-like networks has a continuous nature (see Fig.~S5) similarly to classical percolation transitions. Thus, corruption processes seem to operate close to a critical recidivism rate below which the network becomes entirely fragmented and above which it is overly connected.

To compare our model with the empirical results, we have simulated an ensemble of one hundred networks using the recidivism rate of each country while fixing all other parameters ($s_c=7$, $\beta = 12$, and $p = 0.025$). In these simulations, the number of complete graphs added to the networks is set equal to the total number of scandals in each dataset (437 for Spain and 65 Brazil). Besides generating very similar networks to their empirical counterparts (Fig.~S1), our model approximately replicates several structural properties of real corruption networks. When comparing the simulated results (average value $\pm$ standard deviation over the ensemble) with the Spanish network, we find very similar clustering ($0.949 \pm 0.003$ vs $0.908$ for the entire network and $0.945 \pm 0.004$ vs $0.939$ for the giant component) and assortativity coefficients ($0.76 \pm 0.01$ vs $0.74$ for the entire network and $0.69 \pm 0.02$ vs $0.59$ for the giant component), and somewhat comparable values for density ($0.0042 \pm 0.0002$ vs $0.007$ for the entire network and $0.0084 \pm 0.0009$ vs $0.025$ for the giant component) and the average shortest path ($9.17 \pm 1.01$ vs $5.11$ for the giant component). For the Brazilian network, the model yields similar clustering coefficients ($0.938 \pm 0.008$ vs $0.925$ for the entire network and $0.938 \pm 0.007$ vs $0.929$ for the giant component), assortativity coefficients ($0.67 \pm 0.03$ vs $0.53$ for the entire network and $0.63 \pm 0.04$ vs $0.50$ for the giant component), and densities ($0.030 \pm 0.003$ vs $0.044$ for the entire network and $0.043 \pm 0.008$ vs $0.067$ for the giant component), and somewhat comparable average shortest paths ($4.87 \pm 0.55$ vs $2.99$ for the giant component). The degree distributions of the simulated networks are also in good agreement with exponential distributions, but with characteristic degrees somewhat smaller than the empirical ones ($14.00 \pm 0.66$ vs $20.0$ for Spain and $14.84 \pm 0.72$ vs 17.6 for Brazil).

In addition to static properties, we have also verified that our model replicates the growth process of corruption networks. In particular, we find that the degree distributions of simulated networks are well-described by exponential distributions with characteristic degrees approaching a constant value for later network stages (Fig.~S2). The evolution of simulated networks is also marked by the coalescence of components, which in turn reproduces the abrupt changes observed in the size of the largest and second-largest components (Fig.~S3). Simulated networks also display modular structures that often merge two or more scandals into single network modules. Moreover, the association between the number of network modules and scandals is linear over the entire growth of simulated networks (Fig.~S4), although the average ratios between number of modules and scandals estimated from simulations with the recidivism rates of Spain and Brazil ($0.813\pm0.001$ and $0.783\pm0.001$, respectively) are slightly larger than the empirical ones ($0.744\pm0.004$ and $0.63\pm0.02$, respectively).

While the agreement between properties of empirical and simulated networks is far from perfect, it remains surprising that such a simple model qualitatively replicates all features of our corruption networks, including dynamical properties. Part of the discrepancies between data and model (such as the smaller characteristic degree and the larger average shortest paths obtained in the simulations) can be attributed to the deviations observed between the exponential distribution and the size distribution of scandals (Fig.~\ref{fig:1}). However, these deviations also indicate that other processes are likely to affect the structure of corruption networks. An exciting possibility that future investigations can explore has to do with the fact that our model does not distinguish among corrupt agents. This distinction is crucial in the present context of rising political polarization~\cite{campbell2018polarized, stewart2020polarization, leonard2021nonlinear, waller2021quantifying}, where one may expect partisan and ideological divisions to be also reflected in political corruption and thus on the structure of corruption networks. Besides the likely importance of this and other mechanisms related to corruption processes, our findings indicate that the recidivism of a small fraction of corrupt agents is crucial for the structure and dynamics of corruption networks.

\section*{Discussion}

We have presented an extensive characterization of static and dynamical properties of corruption networks related to political scandals in Spain and Brazil. Despite important differences in the political systems of both countries, our results have shown that the Spanish and Brazilian corruption networks share surprisingly similar structural and dynamical properties. This universality indicates that corruption processes share universal features that are independent of social and cultural differences among countries, as well as independent of individual psychological attributes of corrupt agents. Moreover, we have proposed a simple model in which the recidivism rate is the main ingredient to strengthen this hypothesis. Simulations of our model not only qualitatively replicate all properties of the empirical networks but also indicate that corruption processes appear to operate near a critical recidivism rate. Corruption networks simulated below this critical recidivism rate are completely fragmented, while networks generated above this critical value become overly connected. 

Taken together, empirical results and simulations indicate that a few recidivist agents typically play a prominent role in corruption activities. These agents act as bridges among minor corrupt groups and possibly engage and coordinate them to work in more extensive and often much more harmful corruption processes to society. Considering the many adverse impacts of corruption on democracy~\cite{transparencyhow}, economy~\cite{mauro1995corruption, shao2007quantitative}, and on the trust in the rule of law~\cite{corruptionsundermines}, our findings indicate that public policies and operational law enforcement activities focused on decreasing corruption recidivism, such as increasing the severity of sentences, swift legal processes, and strict serve of sentences, are likely to have a significant negative impact on this type of organized crime by reducing the overall connectivity of corruption networks.

However, since our results are based on corruption scandals of two western countries, and despite the difficulties in finding information about corruption processes, future work should be, if at all possible, dedicated to other countries in order to further strengthen or limit the universalities that we report. Moreover, the lack of quantitative agreement between our model and some empirical properties of corruption networks suggests that factors other than recidivism may affect the structure of political corruption networks. These factors may include political polarization, demography, agent adaptation, and memory effects. There is thus certainly room for the development of other, likely more complex, network models to describe organized crime. Another limitation of our work concerns the information quality used to create corruption networks. Despite the best efforts to make these data reliable, as it happens with all data related to illegal activities, ours may suffer from two types of bias. First, being named in a corruption scandal does not guarantee that a particular person has been convicted of a crime or done anything illegal. Second, it is likely that some people involved in corruption scandals have not been identified during investigations. The compilation of data on corruption will always suffer, at least to a certain degree, from such limitations. Still, we have found that our empirical findings are very robust against randomly removing a fraction of scandals from our data set (see Figs~S6-S9), indicating that the general patterns of corruption processes uncovered by our work are not affected by such biases. Thus, and despite these limitations, we believe that our work contributes significantly to better understand organized crime as a complex networked system, and to identify the essential features of corruption networks that may lead to better criminal policies and more efficient law enforcement interventions.

\bibliography{references.bib}

\section*{Acknowledgements}
We acknowledge the support of the Coordena\c{c}\~ao de Aperfei\c{c}oamento de Pessoal de N\'ivel Superior (CAPES -- PROCAD-SPCF Grant 88881.516220/2020-01), the Conselho Nacional de Desenvolvimento Cient\'ifico e Tecnol\'ogico (CNPq -- Grants 407690/2018-2, 303121/2018-1, 304634/2020-4, and 303533/2021-8), and the Slovenian Research Agency (Grants J1-2457 and P1-0403).

\section*{Data availability}
The datasets used during the current study are freely available as a supplementary file in Ref.~\cite{ribeiro2018dynamical} (the Brazilian corruption network) and can be download from the web page \url{casos-aislados.com} (the Spanish corruption network).

\section*{Author contributions statement}
A.F.M., B.R.d.C., Q.S.H., S.G., M.P., and H.V.R. designed research, performed research, analyzed data, and wrote the paper.

\clearpage
\includepdf[pages=1-9,pagecommand={\thispagestyle{empty}}]{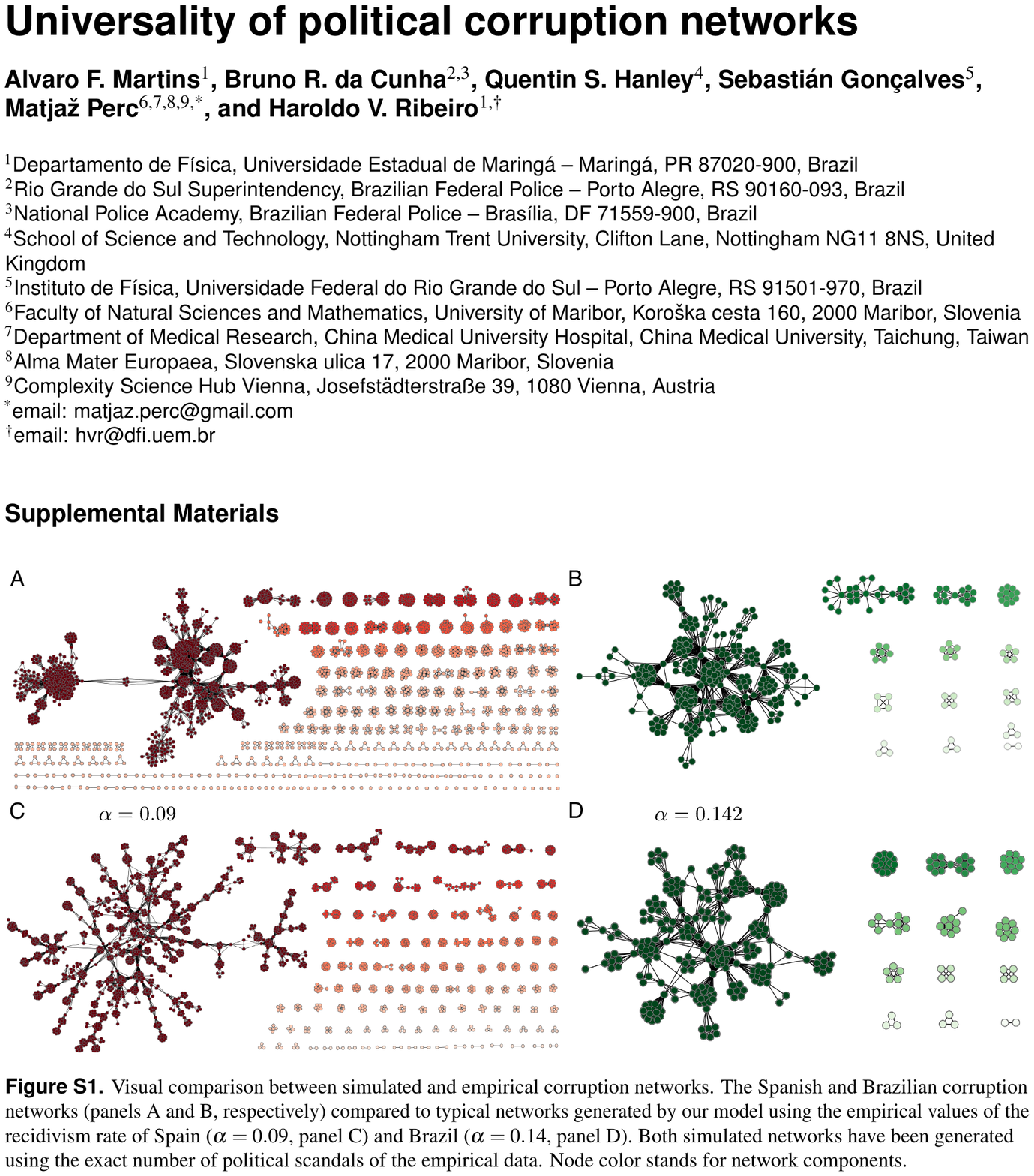}

\end{document}